\documentclass[twoside]{article}
\usepackage{fleqn,espcrc2,epsf}

\usepackage{graphicx}
\newcommand{\pslash}[1]{\rlap{/}\kern-0.8pt #1}
\newcommand{\lslash}{\rlap{/}\kern-0.0pt l}
\newcommand{\Dslash}{\rlap{/}\kern-2.0pt D}

\newcommand{\AmS}{{\protect\the\textfont2
  A\kern-.1667em\lower.5ex\hbox{M}\kern-.125emS}}

\title{Domain Wall Fermion Study of Scaling in Non-perturbative 
Renormalization of Quark Composite Operators}

\author{Y. Zhestkov\address[CU]{Physics Department,
		Columbia University,
		New York, NY 10027}
\thanks{
The author thanks C.~Dawson for much important assistance with
the NPR technique and N.~Christ, T.~Blum, and R.~Mawhinney
for helpful discussions.
This work was supported in part by the U.S. Department of Energy and
the RIKEN BNL Research Center.
}
}

\begin{document}

\begin{abstract}
We compute non--perturbatively the renormalization coefficients
of scalar and pseudoscalar operators, local vector and axial currents,
conserved vector and axial currents, and $O^{\Delta S=2}_{LL}$
over a wide range of energy
scales using a scaling technique that connects the results of
simulations at different values of coupling $\beta$. We use the domain
wall fermion formulation in the quenched approximation at a series of
three values of $\beta = 6.0, 6.45, 7.05$ corresponding to lattice
spacing scaling by factors of two.  
\vspace{1pc}
\end{abstract}

\maketitle

\section{INTRODUCTION}

In lattice QCD
a method for full non--perturbative renormalization of operators
by means of Monte--Carlo simulations was proposed in \cite{Mart95}.
To extend the range of momenta over which operators are renormalized,
we combine simulations at different values of 
coupling constant $\beta$ but with
overlapping regions of physical momenta,
relating renormalization coefficients between momenta scales from $2$ to $10$
GeV.

\section{NONPERTURBATIVE RENORMALIZATION METHOD}

The renormalized operator $O(\mu)$ is defined as
\begin{equation}
      O(\mu) = Z_O(\mu; a) O_{bare}(a)
\end{equation}
where for an operator with $n$ quark fields $Z_O(\mu;a)$ 
is defined by the renormalization condition
\begin{equation}\label{rencond}
  Z_O(\mu; a) Z_{\psi}^{-n/2}(\mu; a) \Gamma_O(p;a) |_{p^2=\mu^2} = 1 \;.
\end{equation}
Here $\Gamma_O(p)$ is a truncated Greens function traced with
a projector $P_O$ on the tree--level operator. It is calculated 
with far off--shell external momenta in Landau gauge.
For quark bilinears, 
$\Gamma_O(p;a) = {\mbox{Tr}\left(\Lambda_O(p;a) \hat{P}_O\right)/12}$,
where $\Lambda_O(p;a) = S(p;a)^{-1} G_O(p;a) S(p;a)^{-1}$,
and $G_O(p;a)$ is non--amputated Greens function.

The quark field renormalization coefficient is defined through
the Ward identity for the conserved vector current,
\begin{equation}\label{Zpsi}
  Z_\psi(\mu;a) = {1\over 48} \mbox{Tr}\left(\gamma^\mu 
                \Lambda^\mu_{V^C}(p;a)\right)_{p^2=\mu^2} \;.
\end{equation}

\section{SCALING TECHNIQUE}

When scaling violations are small,
the renormalization coefficients calculated at
two different values of the lattice spacing are related to each
other by an overall multiplicative coefficient which is independent of
the physical conditions such as physical volume $V$ or momentum scale $\mu$,

\begin {equation}\label{rescaling}
  {Z_O(\mu, V; a')\over Z_O(\mu, V; a)} = R(a', a)
\end{equation}

For each operator we perform a series of simulations with
lattice spacing $a$ increasing by factors of two. The number of lattice
sites in all directions is fixed. The physical volume is finite and
different in each simulation, increasing by factors of $2^4$.

We rescale the $Z$--factors using property (\ref{rescaling})
so that they are all defined at a single value of $\beta=7.05$.
To compute the scaling coefficients $R(a',a)$,
a series of additional simulations is used with half the number of
the lattice sites in each direction and twice as large lattice spacings,
so that
the physical volume and momenta match those in the main series.
We use fits quadratic in $\mu a$ to remove the discretization
error corrections to (\ref{rescaling}) that for domain wall fermions
are of order $(\mu a)^2$,
\begin{equation}\label{ratioaOp2}
 R(a,2a) = {Z_O(\mu,V;a)\over{Z_O(\mu,V;2a)}} + O\left((\mu a)^2\right) \;.
\end{equation}
Once we know the rescaling coefficients for pairs of
lattice spacings, $R(a,2a)$, $R(2a, 4a)$, etc., we use them to rescale
the $Z$--factors obtained at different $\beta$'s,
\begin {equation}
  %Z_O(\mu,V;2a) \rightarrow 
Z_O(\mu,V;a)=R(a,2a) Z_O(\mu,V;2a)
\end {equation}
\begin {equation}
  %Z_O(\mu,V;4a) \rightarrow 
Z_O(\mu,V;a)=R(a,2a)R(2a,4a) Z_O(\mu,V;4a)
\end{equation}
etc.

An important advantage of this method is that the physical volume
is explicitly finite in the simulations at intermediate $\beta$'s.
The infinite volume limit can be taken at the largest
$\beta$ to match to perturbation theory.
\begin{figure}[htb]
%\epsfxsize=\hsize
%\vspace{-20pt}
\epsfbox{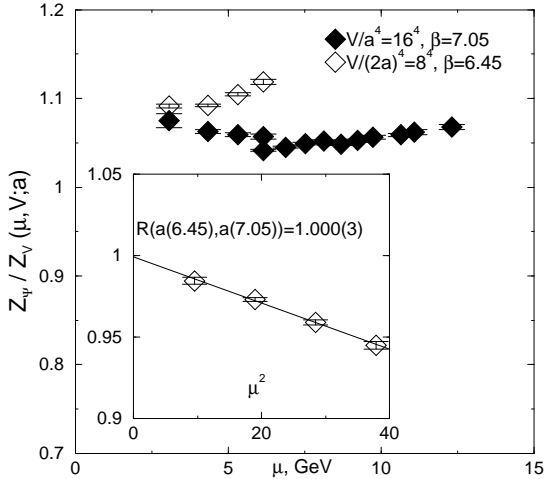}
\vspace{-30pt}
\caption{$Z_\psi/Z_V^L$ in the chiral limit
for the local vector current in the same physical volume
and rescaling coefficient $R(a(6.45),a(7.05))$.}
\label{fig:Zv}
\vspace{-18pt}
\end{figure}
\begin{figure}[htb]
%\epsfxsize=\hsize
%\vspace{-20pt}
\epsfbox{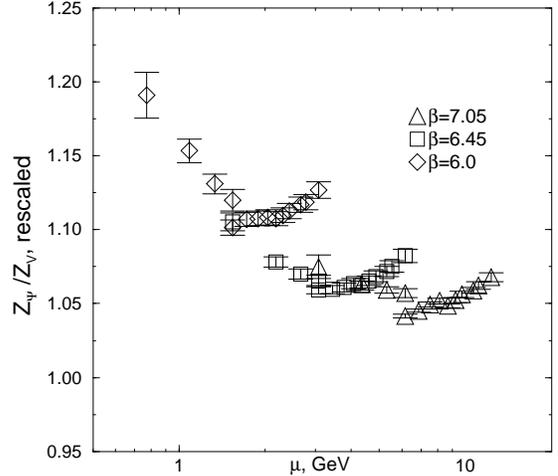}
\vspace{-35pt}
\caption{$Z_\psi/Z_V^L$ at different $\beta$'s, rescaled
to $\beta=7.05$.}
% Solid line -- RG running of $Z_\psi$.}
\label{fig:Zvall}
\vspace{-20pt}
\end{figure}
\section{NUMERICAL RESULTS}
In our simulations we use
lattice volumes $8^4$ and $16^4$ at three values of $\beta=6.0$,
$6.45$, and $7.05$ with the inverse lattice spacing
of $1.96$, $3.92$, and $7.84$ GeV respectively \cite{Bali93}.
\subsection{Local Vector and Axial Currents}
Figure \ref{fig:Zv} illustrates the procedure of obtaining
the rescaling coefficient for the local vector current
from two simulations at different $\beta$'s  in the same physical volume.
The two sets of symbols on the main graph correspond to $Z_\psi/Z_V(\mu,V;a)$
for two values of 
the lattice spacing at $\beta=7.05$ and $6.45$
different by a factor of two. The smaller plot shows a linear extrapolation
of the ratio of these $Z$--factors to $\mu=0$ according to (\ref{ratioaOp2}).

\begin{figure}[htb]
\vspace{-14pt}
\epsfbox{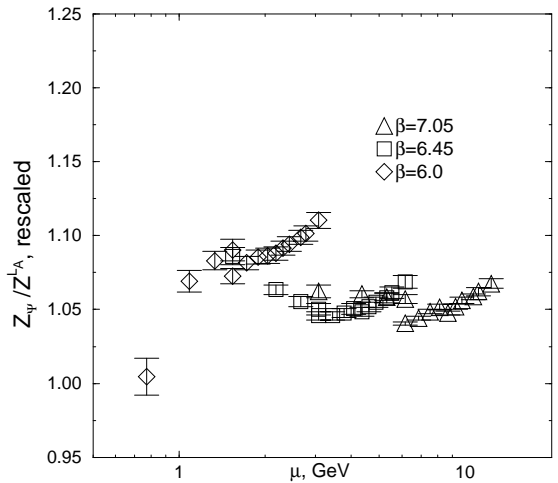}
\vspace{-34pt}
\caption{$Z_\psi/Z_A^L$ at different $\beta$'s, rescaled
to $\beta=7.05$.}
% Solid line -- RG running of $Z_\psi$.}
\label{fig:Zaall}
\vspace{-18pt}
\end{figure}

Figures \ref{fig:Zvall} and \ref{fig:Zaall} show rescaled $Z$--factors
defined at $\beta=7.05$. 
%The solid line is the renormalization group 
%running of $Z_{\psi}$.
\begin{figure}[htb]
%\epsfxsize=\hsize
%\vspace{-20pt}
\epsfbox{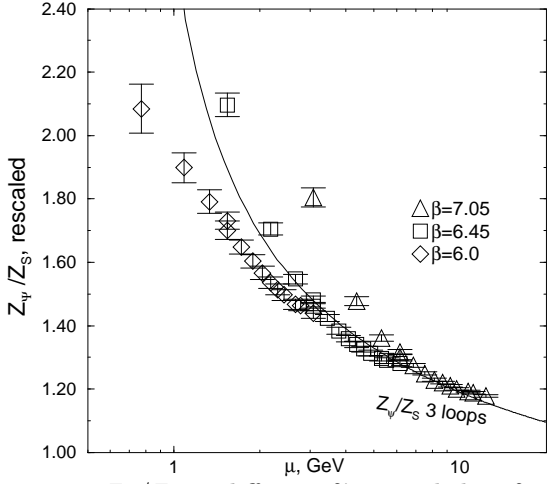}
\vspace{-35pt}
\caption{$Z_\psi/Z_S$ at different $\beta$'s, rescaled
to $\beta=7.05$. Solid line -- RG running of $Z_\psi/Z_S$.}
\label{fig:Zsall}
\vspace{-18pt}
\end{figure}

\subsection{Scalar and Pseudoscalar Densities}

\begin{figure}[htb]
%\epsfxsize=\hsize
%\vspace{-25pt}
\epsfbox{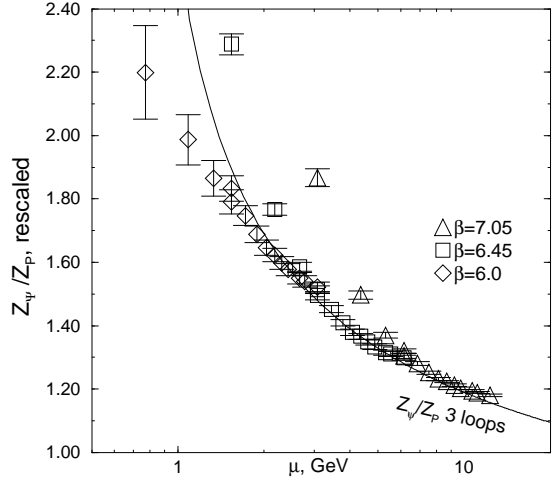}
\vspace{-30pt}
\caption{$Z_\psi/Z_P$ at different $\beta$'s, rescaled
to $\beta=7.05$. Solid line -- RG running of $Z_\psi/Z_P$.}
\label{fig:Zpall}
\vspace{-18pt}
\end{figure}

When taking the chiral limit of scalar and pseudoscalar densities
 at energy scales of a few GeV in finite
volume, one needs to 
remove non--leading $1/p^2$ contributions which have
a singular behavior as $m\rightarrow 0$. These $1/p^2$ contributions
are proportional to the chiral condensate,
\begin{equation}
   \Gamma_P(p^2) = {Z_{\psi}\over Z_P}(p^2) + {C\over p^2}
        {\langle \bar{\psi}\psi\rangle\over m_f} \;.
\end{equation}
\begin{equation}\label{222}
 \Gamma_S(p^2) = {Z_{\psi}\over Z_S}(p^2) + {C\over p^2}
        {\partial\langle\bar\psi\psi\rangle\over\partial m_f} \;.
\end{equation}
For very small masses the zero modes make a significant contribution to 
$\langle\bar\psi\psi\rangle$ with a term inversely proportional to
$m_f$. This term contributes to (\ref{222}) as $-C'/(m_f^2 p^2)$.
After subtracting these terms, the scaling procedure is applied with
the results shown in Figures \ref{fig:Zsall} and \ref{fig:Zpall}.
%%%%%%%%%%%%%%%%%%%
\subsection{Conserved Vector and Axial Currents}
The conserved currents can be used to compute $Z_\psi$ from (\ref{Zpsi}).
With the DWF action,
\begin{equation}
J_\mu(x) = \left(\sum_{s=L_s/2}^{L_s-1}
 + \sigma_J \sum_{s=0}^{L_s/2-1} \right)j_\mu(x,s)\;,
\end{equation}
\begin{eqnarray}\nonumber
j_\mu(x,s) = 
      \bar\psi_s(x)\frac{1-\gamma_\mu}{2}U_\mu(x)\psi_s(x+\hat\mu)\\
   - \bar\psi_s(x+\hat\mu)\frac{1+\gamma_\mu}{2}U^\dagger_\mu(x)\psi_s(x) \;,
\end{eqnarray}
where $\sigma_V=+1$ and $\sigma_A=-1$.

Due to the nonlocal character and summation over $s$, these matrix elements
can be very expensive to calculate. To reduce the amount of computational
time, we used a random source estimator to compute the part of the sum between
$s=1$ and $L_s-2$ and 
used only one component of $\Gamma^\mu(p)$ for
different momenta to calculate
\begin{eqnarray}
\nonumber
  \mbox{Tr}\left(\gamma^\mu \Gamma^\mu(p)\right)=
   \mbox{Tr}\{\gamma^0\Gamma^0(p)
+ \gamma^0\Gamma^0(p^0\leftrightarrow{p^1})\\
+ \gamma^0\Gamma^0(p^0\leftrightarrow{p^2})
+ \gamma^0\Gamma^0(p^0\leftrightarrow{p^3}) \}\;.
\end{eqnarray}
\begin{figure}[htb]
\vspace{-20pt}
\epsfbox{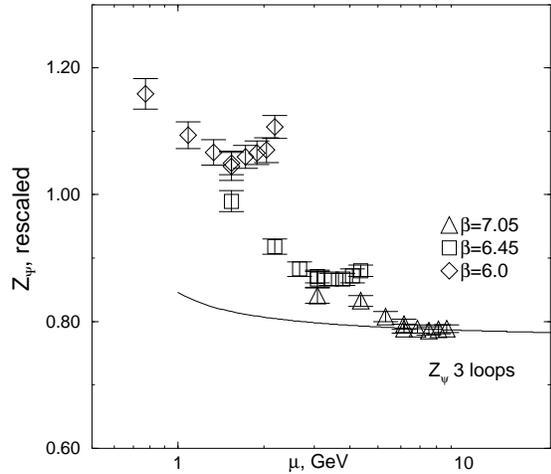}
\vspace{-35pt}
\caption{$Z_\psi$ from conserved vector current, rescaled
to $\beta=7.05$. Solid line -- RG running.}
\label{fig:Vectall}
\vspace{-30pt}
\end{figure}
\begin{figure}[htb]
%\epsfxsize=\hsize
%\vspace{-20pt}
\epsfbox{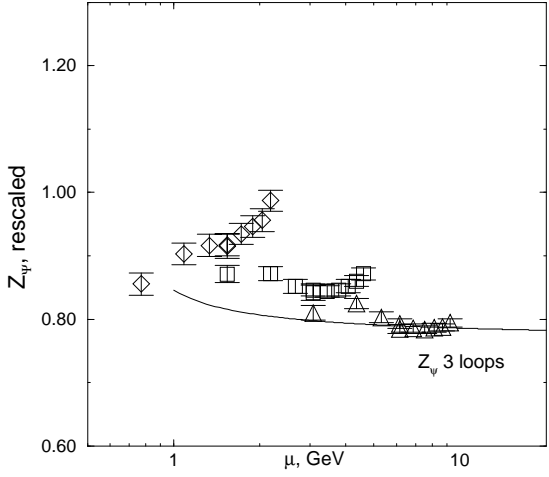}
\vspace{-30pt}
\caption{$Z_\psi$ from conserved axial current, rescaled
to $\beta=7.05$. Solid line -- RG running.}
\label{fig:Axall}
\vspace{-18pt}
\end{figure}
\begin{figure}[htb]
%\epsfxsize=\hsize
%\vspace{-20pt}
\epsfbox{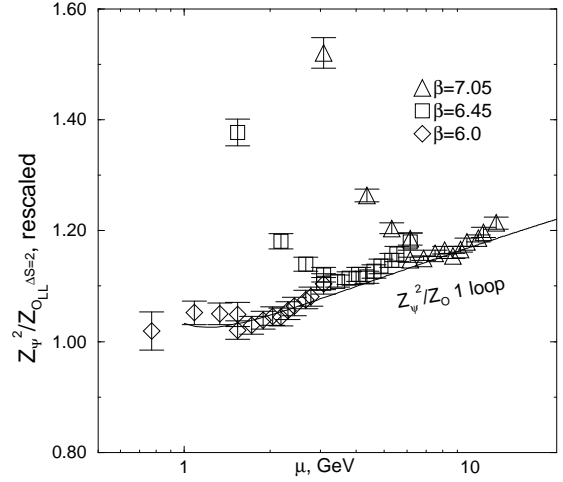}
\vspace{-30pt}
\caption{$Z_\psi^2/Z_{O_{LL}^{\Delta S=2}}$ at different $\beta$'s, rescaled
to $\beta=7.05$. Solid line -- RG running.}
\label{fig:Zbkall}
\vspace{-18pt}
\end{figure}
We set $m_f=0.05$ in these calculations.
Figures \ref{fig:Vectall} and \ref{fig:Axall} show rescaled $Z_\psi$
defined at $\beta=7.05$ obtained from the conserved vector and axial
currents.
% Solid line is the renormalization group running of $Z_\psi$.
%%%%%%%%%%%%%%%%%%%
\subsection{$B_K$}
$B_K$ is a phenomenologically important parameter relevant to the
calculation of $K^0-\bar{K}^0$ mixing,
\begin{equation}
 B_K(\mu) = {\langle\bar{K}^0|{O}_{LL}^{\Delta{S}=2}(\mu)|K^0\rangle
  \over
\frac{8}{3}f^2_K m^2_K} \;.
\end{equation}
The renormalization of $B_K$ is the same as that of
$O_{LL}^{\Delta S=2}=(\bar{s}\gamma^\mu_L{d})^2$.
Using the upper indices for color and the lower for spin, the
non--amputated Fourier transformed four--point Greens function with
equal off-shell external momenta can be written as
\begin{eqnarray}\nonumber
  G_{B_K}(p)^{ABCD}_{\alpha\beta\gamma\delta} =
   2[ 
    \langle\Gamma^\mu(p)^{AB}_{\alpha\beta}  
        \Gamma^\mu(p)^{CD}_{\gamma\delta} \rangle\\
    -  \langle\Gamma^\mu(p)^{AD}_{\alpha\delta}  
        \Gamma^\mu(p)^{CB}_{\gamma\beta} \rangle
   ] \;,
\end{eqnarray}
where $\Gamma^\mu(p)^{AB}_{\alpha\beta} = S(p|0)^{AR}_{\alpha\sigma}
         {\gamma^\mu_L}_{\sigma\rho}
         \left(\gamma_5 S(p|0)^\dag \gamma_5\right)^{RB}_{\rho\beta}$.
$S(p|0)$ is a propagator on a single configuration. 
The projection operation is defined as
\begin{equation}
\Gamma_{B_K}(p;a) = 
    {{\gamma^\mu_L}_{\alpha\alpha'} {\gamma^\mu_L}_{\beta\beta'}
    \Lambda_{B_K}(p)^{AABB}_{\alpha'\alpha\beta'\beta}
\over 32N_c(N_c+1)} \;,
\end{equation}
with the amputated Greens function given by
\begin{eqnarray}\nonumber
\Lambda_{B_K}(p)^{ABCD}_{\alpha\beta\gamma\delta}
  = S^{-1}(p)^{AA'}_{\alpha\alpha'} S^{-1}(p)^{CC'}_{\gamma\gamma'}
\;\;\;\;\;\;\;\;\;\\
     \quad \cdot \;G_{B_K}(p)^{A'B'C'D'}_{\alpha'\beta'\gamma'\delta'}
   S^{-1}(p)^{B'B}_{\beta'\beta} S^{-1}(p)^{D'D}_{\delta'\delta} \;.
\end{eqnarray}
Applying our earlier analysis then gives the results shown in Figure
\ref{fig:Zbkall}. 

We conclude that this combination of recursive
scaling and NPR offers a promising connection between the perturbative
and non--perturbative regimes--a connection that is required if
perturbative errors are to be properly controlled.  The results
reported here suggest a consistent picture for the scaling of 
renormalization factors over a momentum range between 2 and 10 GeV for
the case of quenched domain wall fermions.  More extensive
calculations will be required for a complete demonstration of this
approach.

\end{document}